\documentclass[12pt,aps,showpacs,showkeys]{revtex4}
\usepackage{hyperref}
\usepackage{amsfonts}
\usepackage{amsmath}
\usepackage{amssymb}
\usepackage{graphicx}

\def\Tr{{\rm Tr}}

\begin{document}

\title{Multi-particle and High-dimension Controlled Order Rearrangement Encryption Protocols}

\author{Ya Cao, An-Min Wang, Xiao-San Ma and Ning-bo Zhao }
\email{caoy1209@mail.ustc.edu.cn}
 \affiliation{Department of Modern Physics,
University of Science and Technology of China, Hefei 230026, China}

\keywords{ quantum cryptography, ~quantum key distribution,
~controlled order rearrangement encryption, ~maximally entangled
state, ~covariant quantum cloning machine }

\pacs{03.67.Dd, 03.67.Hk}
\begin{abstract}
Based on the controlled order rearrange encryption (CORE) for
quantum key distribution using EPR pairs[Fu.G.Deng and G.L.Long
Phys.Rev.A68 (2003) 042315], we propose the generalized controlled
order rearrangement encryption (GCORE) protocols of $N$ qubits and
$N$ qutrits, concretely display them in the cases using 3-qubit,
2-qutrit maximally entangled basis states. We further indicate that
our protocols will become safer with the increase of number of
particles and dimensions. Moreover, we carry out the security
analysis using quantum covariant cloning machine for the protocol
using qutrits. Although the applications of the generalized scheme
need to be further studied, the GCORE has many distinct features
such as great capacity and high efficiency.

\end{abstract}

\maketitle
\section{Introduction}

Cryptography is an art of providing secure communication over
insecure communication channels. Now, in the information community,
the safety of transmission of secret information is getting more and
more important. One essential theme of secure communication is to
distribute secret keys between sender and receiver. Quantum
cryptography (QC) is secure based on the fundamental principles of
quantum mechanics rather than classical cryptography. An important
application of QC is the quantum key distribution (QKD), which
concerns the generation and distribution of secret key between two
legitimate users. The security of  key distribution is the most
important part in the secret communication. QKD exploits quantum
mechanics principles for the secret communication, which provides a
secure way for transmitting the key. So far, there are many quantum
secret key protocols such as BB84 protocol, Ekt91, B92, six-state
protocol etc.[1-5], and some new quantum secret key protocols [6-12]
are continually suggested.

The security of some QKD protocols in Refs.[1,2,3,6,10-14] are based
on random choices of different measuring-base, so the randomness is
usually a useful ingredient in QC. The security of other some QKD
protocols in Refs.[7,8,15-17] lies on the nonlocality nature of
quantum systems. Goldenberg-Vaidman scheme [7] first presents QKD
protocol by two transmission lines. This protocol uses orthogonal
states and has full efficiency, all the particles transmitted are
used to generate secret keys. Then, Koashi-Imoto protocol [8]
improves Goldenberg-Vaidman scheme by using an asymmetric
interferometer to reduce the time delay. However, two factors lead
that these schemes' time delay can not be too short. Subsequently,
F.G.Deng and G.L.Long propose a controlled order rearrange
encryption (CORE) scheme [9] to overcome this drawback and realize a
secure QKD. In the nonlocality based QKD protocols, orthogonal
quantum states are used. Security is assured by not allowing an
eavesdropper such as Eve to acquire both parts simultaneously.

Actually, the CORE technique is implemented not only suitable to use
Einstein-Podolsky-Rosen (EPR) pairs, but also suitable to use other
quantum information carriers (QICs) [9]. In recent years researchers
have drawn their attentions to the QKD protocols that involve
multilevel systems with two parties, or multiple parties with
two-level systems. A pursuit motivation of multilevel QKD is that
more information can be carried by each particle thereby the
information flux is increased, and some multilevel protocols have
been shown to have greater security against eavesdropping attacks
than their qubit-based countparts [16,18,19]. Thus, the use of
multi-particle maximally entangled state can guarantee the security
further and has higher efficiency in general.

In this paper, our main purpose is to generalize the CORE of QC to
multi-particle and/or high dimension quantum systems. Our
generalized protocols can be thought of having higher efficiency
because the generalized protocols, which is called as the GCORE of
QC here, exploit the facts that a possible eavesdropper with no
access to the whole quantum system at the same time, cannot
recover the whole information without being detected, and the
protocols employ a larger alphabet, a few-dimensional orthogonal
basis of pure state. Consequently, we obtain the full efficiency
from this point of view. The generalized protocols also have great
capacity based on the reason that $M$ adopted $N$- qudit maximally
entangled states can send $M\log _2 d^N$ bits of information in
our schemes if we assume there are $N$ particles with each being
$d$ dimension.

The paper is organized as follows. In Sec.II, we simply review the
CORE protocol using EPR pairs provided by Fu.G.Deng and G.L.Long.
Then we generalize the CORE protocol to $N$-qubit case, specially,
we present the GCORE protocol using 3-qubit state and check its
security by the correlated matrix method. In Sec.III, the GCORE
protocol using $N$ qutrits is proposed, GCORE protocol using
2-qutrit is presented in detail. Moveover, we discuss the security
of qutrit GCORE using the quantum covariant cloning machine. In
Sec.IV, we present a uniform expression of multi-particle and/or
high dimension situation. Advantages of GCORE are analogized and
concluding remarks are given.

\section{GCORE using $N$-qubit maximally entangled basis states}\label{sec1}

\subsection{Explanation of CORE protocol}
At the beginning, let us review briefly the meaning of CORE. Assume
the keys are distributed between Alice and Bob. Before transmission,
Alice rearranges the order of correlated particles and sends them to
Bob. The aim of random rearrangement is to prevent the eavesdropper
obtaining correlated particles simultaneously from different
transmission channels as possible as they can, and we also need an
evening process to make transmission in equal time intervals. Once
Bob receives these particles, he restores the order of the particles
and undoes Alice's operations by synchronizing their measure devices
using repeatedly a {\it prior} shared control key, so that he can
make orthogonal basis measurement. The measurement outcome is
exactly what Alice has prepared. The essence of CORE is use of a a
control key as has been used in the modified BB84 scheme [6]. The
noncloning nature ensures it viable.

The whole process of CORE protocol using EPR states [9] has been
demonstrated clearly in Ref.[9]. In the following contexts, we
generalize it to multi-particle and high-dimensional cases, hence
the generalized protocol is denoted as GCORE.

\subsection{GCORE protocol using GHZ-basis states}
In the following, we firstly discuss concrete GCORE example using
3-qubit GHZ-basis states without loss of generalization.

(i) Alice generates a sequence of GHZ-basis states
$(a_1,b_1,c_1),\cdots, (a_m,b_m,c_m)$ randomly, where
$(a_i,b_i,c_i)$ denotes one GHZ-basis state ($1 \leq i \leq m, m$
is an integer) and every eight adjoining triplets are taken as one
unit of QICs. Without loss of generality, we consider the first
carrier unit [$(a_1,b_1,c_1),(a_2,b_2,c_2), \cdots,
(a_8,b_8,c_8)$] which are randomly in eight GHZ-basis states that
can be expressed as [20]:

\begin{equation}
\left| {\psi _j^\pm } \right\rangle = \frac{1}{\sqrt 2 }\left(
{\left| j \right\rangle _{AB} \left| 0 \right\rangle_C \pm \left| {3
- j} \right\rangle _{AB} \left| 1 \right\rangle _C } \right),
\end{equation}

\noindent where $j = j_1 j_2 $ denotes binary notations. In their
explicit forms, eight GHZ-basis states reads:
\begin{eqnarray}
&& \left| {\psi _0 ^ + } \right\rangle = \left( {\left|
{000} \right\rangle + \left| {111} \right\rangle } \right) / \sqrt 2  \nonumber\\
&& \left| {\psi _0 ^ - } \right\rangle = \left( {\left|
{000} \right\rangle - \left| {111} \right\rangle } \right) / \sqrt 2  \nonumber\\
&& \left| {\psi _1 ^ + } \right\rangle = \left( {\left|
{010} \right\rangle + \left| {101} \right\rangle } \right) / \sqrt 2  \nonumber\\
&& \left| {\psi _1 ^ - } \right\rangle = \left( {\left|
{010} \right\rangle - \left| {101} \right\rangle } \right) / \sqrt 2  \nonumber\\
&&\left| {\psi _2 ^ + } \right\rangle = \left({\left|
{100} \right\rangle + \left| {011} \right\rangle } \right) / \sqrt 2  \nonumber\\
&&\left| {\psi _2 ^ - } \right\rangle = \left( {\left|
{100} \right\rangle - \left| {011} \right\rangle } \right) / \sqrt 2  \nonumber\\
&& \left| {\psi _3 ^ + } \right\rangle = \left( {\left|
{110} \right\rangle + \left| {001} \right\rangle } \right) / \sqrt 2  \nonumber\\
&& \left| {\psi _3 ^ - } \right\rangle = \left( {\left| {110}
\right\rangle - \left| {001} \right\rangle } \right) / \sqrt 2 )
\end{eqnarray}

\noindent we indicate them by 000, 001, 010, 011, 100, 101, 110, and
111, respectively.

(ii) Alice sends the three parts out in equal time intervals to Bob
through three channels. Before these GHZ-basis states enter into the
insecure transmission channel, their orders are rearranged by the
GCORE system. Here are eight choices of GCORE operations,
corresponding relations are the following:

\[
\begin{array}{l}
 E_0 \leftrightarrow 000,\quad E_1 \leftrightarrow 001,\quad E_2
\leftrightarrow 010,\quad E_3 \leftrightarrow 011  \\
 E_4 \leftrightarrow 100,\quad E_5 \leftrightarrow 101, \quad E_6
\leftrightarrow 110, \quad E_7 \leftrightarrow 111
 \end{array}
\]

\noindent and the GCORE is done for eight GHZ-basis states. Let us
use permutation group notation to express them as following

\[
\begin{array}{l}
 ~~E_0 = \left( \begin{array}{*{20}c}
 1  & 2  & 3  & 4  & 5  & 6  & 7 & 8  \\
 1  & 2  & 3  & 4  & 5  & 6  & 7 & 8  \\
\end{array}  \right) \\
 ~~~~ = \left( 1 \right)\left( 2 \right)\left( 3 \right)\left( 4
\right)\left( 5 \right)\left( 6 \right)\left( 7 \right)\left( 8 \right) \\
 \end{array}£¬
\begin{array}{l}
~~~~~~~E_1 = \left( \begin{array}{*{20}c}
 1  & 2  & 3  & 4  & 5  & 6  & 7 & 8  \\
 2  & 1  & 4  & 3  & 6  & 5  & 8 & 7  \\
\end{array} \right) \\
~~~~~~~~~~~~~~ = \left( \begin{array}{*{20}c}
 1  & 2  \\
\end{array}  \right)\left( \begin{array}{*{20}c}
 3  & 4  \\
\end{array}  \right)\left( \begin{array}{*{20}c}
 5  & 6  \\
\end{array}  \right)\left( \begin{array}{*{20}c}
 7  & 8  \\
\end{array} \right) \\
\end{array}
\]

\[
 \begin{array}{l}
 E_2 = \left( \begin{array}{*{20}c}
 1  & 2  & 3  & 4  & 5  & 6  & 7 & 8  \\
 3  & 4  & 1  & 2  & 7  & 8  & 5 & 6  \\
\end{array}  \right) \\
 ~~~~ = \left( \begin{array}{*{20}c}
 1  & 3  \\
\end{array}  \right)\left( \begin{array}{*{20}c}
 2  & 4  \\
\end{array}  \right)\left( \begin{array}{*{20}c}
 5  & 7  \\
\end{array}  \right)\left( \begin{array}{*{20}c}
 6  & 8  \\
\end{array}  \right) \\
 \end{array}£¬
\begin{array}{l}
 E_3 = \left( \begin{array}{*{20}c}
 1  & 2  & 3  & 4  & 5  & 6  & 7 & 8  \\
 4  & 3  & 2  & 1  & 8  & 7  & 6 & 5  \\
\end{array}  \right) \\
~~~~ = \left( \begin{array}{*{20}c}
 1  & 4  \\
\end{array}  \right)\left( \begin{array}{*{20}c}
 2  & 3  \\
\end{array}  \right)\left( \begin{array}{*{20}c}
 5  & 8  \\
\end{array}  \right)\left( \begin{array}{*{20}c}
 6  & 7  \\
\end{array}  \right) \\
 \end{array}
\]

\[
\begin{array}{l}
 E_4 = \left( \begin{array}{*{20}c}
 1  & 2  & 3  & 4  & 5  & 6  & 7 & 8  \\
 5  & 6  & 7  & 8  & 1  & 2  & 3 & 4  \\
\end{array}  \right) \\
 ~~~~ = \left( \begin{array}{*{20}c}
 1  & 5  \\
\end{array}  \right)\left( \begin{array}{*{20}c}
 2  & 6  \\
\end{array}  \right)\left( \begin{array}{*{20}c}
 3  & 7  \\
\end{array}  \right)\left( \begin{array}{*{20}c}
 4  & 8 \\
\end{array}  \right) \\
 \end{array}£¬
\begin{array}{l}
 E_5 = \left( \begin{array}{*{20}c}
 1  & 2  & 3  & 4  & 5  & 6  & 7 & 8  \\
 8  & 7  & 6  & 5  & 4  & 3  & 2 & 1  \\
\end{array}  \right) \\
~~~~= \left( \begin{array}{*{20}c}
 1  & 8  \\
\end{array}  \right)\left( \begin{array}{*{20}c}
 2  & 7  \\
\end{array}  \right)\left( \begin{array}{*{20}c}
 3  & 6  \\
\end{array}  \right)\left( \begin{array}{*{20}c}
 4  & 5  \\
\end{array}  \right) \\
 \end{array}
\]

\[
\begin{array}{l}
 E_6 = \left( \begin{array}{*{20}c}
 1  & 2  & 3  & 4  & 5  & 6  & 7 & 8  \\
 7  & 8  & 5  & 6  & 3  & 4  & 1 & 2  \\
\end{array}  \right) \\
 ~~~~ = \left( \begin{array}{*{20}c}
 1  & 7  \\
\end{array}  \right)\left( \begin{array}{*{20}c}
 2  & 8  \\
\end{array}  \right)\left( \begin{array}{*{20}c}
 3  & 5  \\
\end{array}  \right)\left( \begin{array}{*{20}c}
 4  & 6  \\
\end{array}  \right) \\
\end{array}£¬
\begin{array}{l}
 E_7 = \left( \begin{array}{*{20}c}
 1  & 2  & 3  & 4  & 5  & 6  & 7 & 8  \\
 6  & 5  & 8  & 7  & 2  & 1  & 4 & 3  \\
\end{array}  \right) \\
~~~~ = \left( \begin{array}{*{20}c}
 1  & 6  \\
\end{array}  \right)\left( \begin{array}{*{20}c}
 2  & 5  \\
\end{array}  \right)\left( \begin{array}{*{20}c}
 3  & 8  \\
\end{array}  \right)\left( \begin{array}{*{20}c}
 4  & 7  \\
\end{array}  \right) \\
 \end{array}
\]
we also show this protocol using Fig.1.

\begin{figure}
\centering
\includegraphics[width=5.00in,height=3.50in]{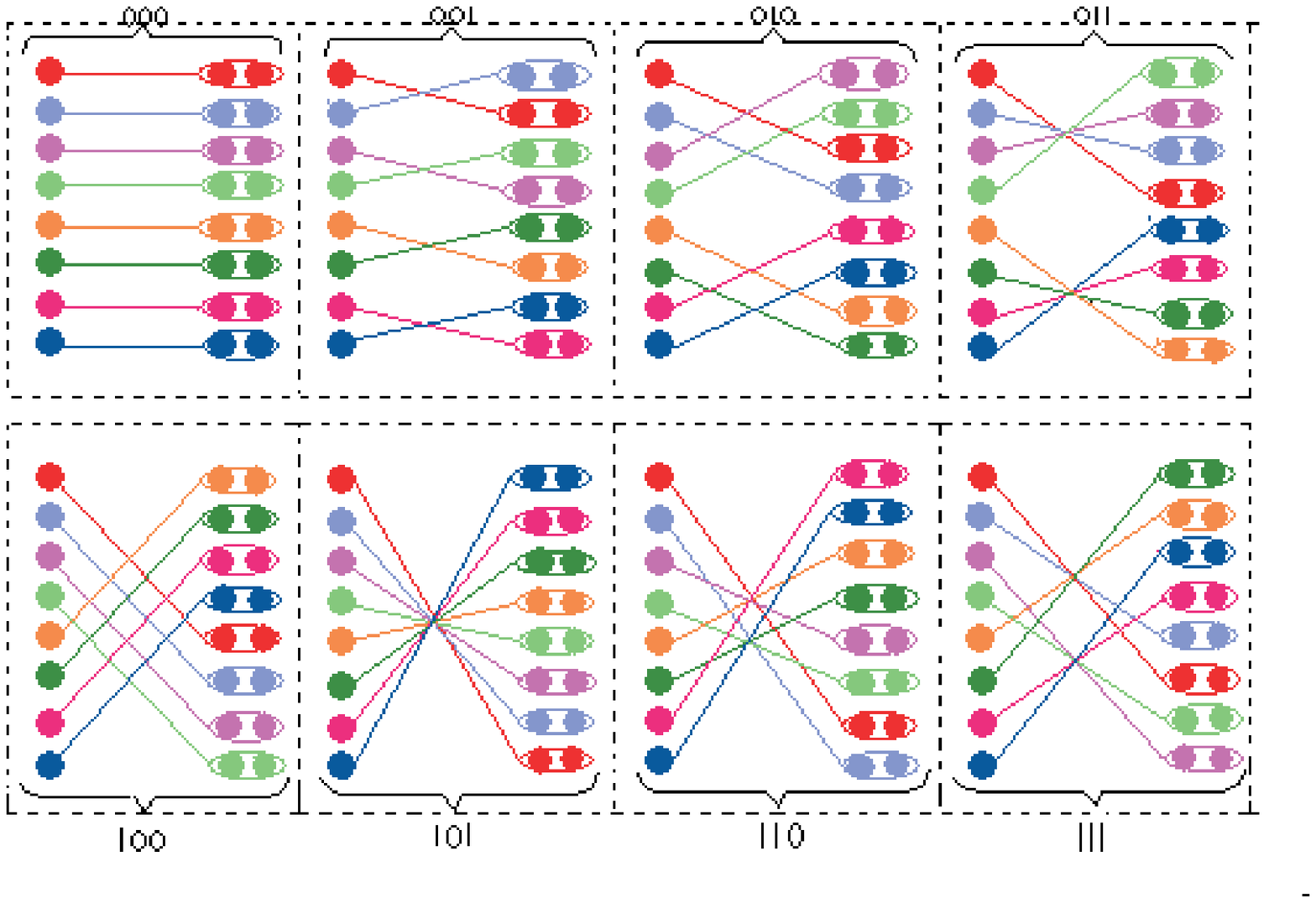}\caption{Example
of GCORE using GHZ-basis states. ~There are eight different GCORE
operations. }
\end{figure}

Three quantum channels in this GCORE protocol are denoted upper,
middle, and lower channel. The upper QIC parts are transmitted
according to their temporal orders. A control key is used to
rearrange the order of middle and lower QIC parts. For instance,
the value of control key is 000, the operation $E_0 $ is applied.
In Fig. 2 there are seven switches, the order of eight GHZ-basis
states is unchanged with switch 1, 2, 3, 4, 5, 6, 7 in position
(up, up, up, up, up, up, down). When the value of control key is
001, the operation $E_1 $ is performed, and it is done by putting
the seven switches into position(down, up, up, up, up, up, down),
(up, up, up, up, up, down,up), (down, up, up, up, up, up, down),
(up, up, up, up, up, down,up), (down, up, up, up, up, up, down),
(up, up, up, up, up, down,up), (down, up, up, up, up, up, down),
(up, up, up, up, up, down,up) for eight particles, respectively.
In fact, five switches are enough. When the operation $E_1 $ is
performed, it is done by putting the five switches into
position(down, down, up, up, down), (up, down, up, down, up),
(down, down, up, up, down), (up, down, up, down, up), (down, down,
up, up, down), (up, down, up, down, up), (down, down, up, up,
down), (up, down, up, down, up) for eight particles, respectively.
The effect of using seven switches is the same as that of using
five switches. Similar combination can be written explicitly for
operations $E_2, E_3, E_4, E_5, E_6, E_7 $.
\begin{figure}[htbp]
\centering
\includegraphics[width=5.40in,height=2.80in]{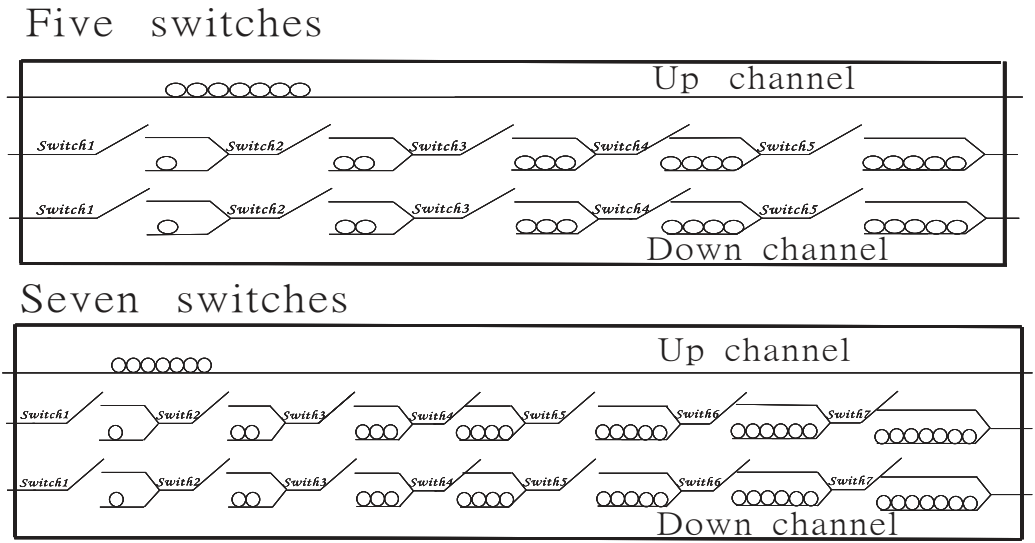}\caption{Devices
to perform GCORE operations, the loop represents a time delay of a
fixed interval.}
\end{figure}

(iii)  Bob undoes the effect of order rearrangement. At Bob' site,
he just exchanges upper, middle, and lower parts of Alice's GCORE
apparatus and the GCORE operations performed by Alice will be
undone. (iv) Bob measures these carrier units to obtain the key.
After these particles are dearranged. Bob uses the GHZ-basis
measurement to read out the information determinatively, which is
exactly the same as Alice prepared one because the measurement
here is orthogonal basis measurement and obviously the eight
GHZ-basis states are mutually orthogonal.

Remark: To prevent Eve from stealing, we need an evening process
to ensure the same time interval between different batches of QICs
travel. Now, we need three transmission lines to ensure the
application of current proposed scheme because 3-qubit GCORE uses
GHZ-basis state, and each particle transmitted through a quantum
transmission line in equal time interval. It is obviously
different from the case using two-transmission lines in
Refs.[7,8]. Detailed analysis will be presented in subsection C
below. In addition, the control keys can be used to control the
GCORE operation of a group of units to reduce resources. For
example, instead of using 001 controls GCORE operation of one unit
of QICs (eight GHZ-basis states), we can use 001 to control more
units of QICs consecutively, say 4 units or 32 GHZ-basis states.

\subsection{Security of GCORE using GHZ-basis states}
Let us look at the security of GCORE using 3-qubit GHZ-basis states.
Eve has only $1/8$ chance to guess the right GCORE operation for the
eight GHZ-basis states. If she uses a wrong GCORE operation, the
three particles measured by her will be anticorrelated. Firstly we
assume that A particle from the first GHZ-basis state, B particle
from the second GHZ-basis state and C particle from the third
GHZ-basis state are mistreated by Eve as a GHZ-basis state, then the
density operator will be
\begin{equation}
 \rho _{A_1B_2C_3} = \tilde {\rho }_{A_1 } \otimes \tilde {\rho
}_{B_2 } \otimes \tilde {\rho }_{C_3 } = \left(
\begin{array}{*{20}c}
 {\displaystyle\frac{1}{2}}  & 0  \\
 0  & {\displaystyle\frac{1}{2}}  \\
\end{array}  \right) \otimes \left( \begin{array}{*{20}c}
 {\displaystyle\frac{1}{2}}  & 0  \\
 0  & {\displaystyle\frac{1}{2}}  \\
\end{array}  \right) \otimes \left( \begin{array}{*{20}c}
 {\displaystyle\frac{1}{2}}  & 0  \\
 0  & {\displaystyle\frac{1}{2}}  \\
\end{array}  \right) = \frac{1}{8}I_{8\times 8}
\end{equation}
\noindent where $\tilde {\rho }_{A_1 } =\Tr_{B_1C_1}\left( {\rho
_{A_1 B_1 C_1 } } \right),\tilde {\rho }_{B_2 } =
\Tr_{A_2C_2}\left( {\rho _{A_2 B_2 C_2 } } \right),\tilde {\rho
}_{C_3 } = \Tr_{A_3B_3}\left( {\rho_{A_3 B_3 C_3 } } \right)$.
When $\rho _{A_1B_2C_3}$ is measured in the GHZ-basis state, the
result can be any one of eight GHZ-basis states with $12.5\%$
probability each. Thus Eve will introduce $66.99\% $ error rate in
the results. Then we assume A particle from the first GHZ-basis
state, B and C particles from the second GHZ-basis state are
mistreated by Eve as a GHZ-basis state, the density operator will
be
\begin{equation}
\rho _{A_1B_2C_3} = \tilde {\rho }_{A_1 } \otimes \tilde {\rho
}_{B_2 C_3 } = \left( \begin{array}{*{20}c}
 {\displaystyle\frac{1}{2}}  & 0  \\
 0  & {\displaystyle\frac{1}{2}}  \\
\end{array}  \right) \otimes \left( \begin{array}{*{20}c}
 {\displaystyle\frac{1}{2}}  & 0  & 0  & 0 \\
 0  & 0  & 0  & 0  \\
 0  & 0  & 0  & 0  \\
 0  & 0  & 0  & {\displaystyle\frac{1}{2}}  \\
\end{array}  \right)
\end{equation}
\noindent Eve will introduce $76.56\% $ error rate in these
results. In both situations, Alice and Bob can detect Eve easily
by checking a sufficiently large subset of results randomly
chosen. Surely, Eve can perform a generalized Bell inequality
measurement on the particles, but it is useless for decrypting the
control key. Let us choose $\vec {a}\left( {a_x ,a_y ,a_z }
\right)$, $\vec {b}\left( {b_x ,b_y ,b_z } \right)$, as the
directions of Alice's and Bob's measurements, at the same time,
$\vec {c}\left( {c_x ,c_y ,c_z } \right)$ is also Bob's
measurement direction. Then the correlation operator can be
written as following:
\begin{equation}
\hat {E} = (\hat {\sigma } \cdot \vec {a} )\otimes (\hat {\sigma }
\cdot \vec {b} )\otimes (\hat {\sigma } \cdot \vec {c})
\end{equation}
\noindent where $\sigma _x = \left( \begin{array}{*{20}c}
 0  & 1  \\
 1  & 0  \\
\end{array}  \right),\sigma _y = \left( \begin{array}{*{20}c}
 0  & { - i}  \\
 i  & 0  \\
\end{array}  \right),\sigma _z = \left( \begin{array}{*{20}c}
 1  & 0  \\
 0  & { - 1}  \\
\end{array}  \right)$.
The expectation values $\left\langle {E\left( {\vec {a},\vec
{b},\vec{c}} \right)} \right\rangle _\psi = \left\langle \psi
\right|(\hat {\sigma } \cdot \vec {a} )\otimes (\hat {\sigma }
\cdot \vec {b}) \otimes (\hat {\sigma } \cdot \vec {c})\left| \psi
\right\rangle $ are different for the different GHZ-basis states.
They are
\begin{eqnarray}
&&\left\langle {E\left( {\vec {a},\vec {b},\vec{c}} \right)}
\right\rangle _{\psi _0^ + } = \left( {a_x - ia_y } \right)\left(
{b_x - ib_y } \right)\left( {c_x - ic_y } \right) + \left( {a_x +
ia_y } \right)\left( {b_x + ib_y } \right)\left( {c_x + ic_y }
\right)\nonumber
\\
&&\left\langle {E\left( {\vec {a},\vec {b},\vec{c}} \right)}
\right\rangle _{\psi _0^ - } = - \left( {a_x - ia_y } \right)\left(
{b_x - ib_y } \right)\left( {c_x - ic_y } \right) - \left( {a_x +
ia_y } \right)\left( {b_x + ib_y } \right)\left( {c_x + ic_y }
\right)\nonumber
\\
&&\left\langle {E\left( {\vec {a},\vec {b},\vec{c}} \right)}
\right\rangle _{\psi _1^ + } = \left( {a_x - ia_y } \right)\left(
{b_x + ib_y } \right)\left( {c_x - ic_y } \right) + \left( {a_x +
ia_y } \right)\left( {b_x - ib_y } \right)\left( {c_x + ic_y }
\right)\nonumber
\\
&&\left\langle {E\left( {\vec {a},\vec {b},\vec{c}} \right)}
\right\rangle _{\psi _1^ - } = - \left( {a_x - ia_y } \right)\left(
{b_x + ib_y } \right)\left( {c_x - ic_y } \right) - \left( {a_x +
ia_y } \right)\left( {b_x - ib_y } \right)\left( {c_x + ic_y }
\right)\nonumber
\\
&&\left\langle {E\left( {\vec {a},\vec {b},\vec{c}} \right)}
\right\rangle _{\psi _2^ + } = \left( {a_x + ia_y } \right)\left(
{b_x - ib_y } \right)\left( {c_x - ic_y } \right) + \left( {a_x -
ia_y } \right)\left( {b_x + ib_y } \right)\left( {c_x + ic_y }
\right)\nonumber
\\
&&\left\langle {E\left( {\vec {a},\vec {b},\vec{c}} \right)}
\right\rangle _{\psi _2^ - } = - \left( {a_x + ia_y } \right)\left(
{b_x - ib_y } \right)\left( {c_x - ic_y } \right) - \left( {a_x -
ia_y } \right)\left( {b_x + ib_y } \right)\left( {c_x + ic_y }
\right)\nonumber
\\
&&\left\langle {E\left( {\vec {a},\vec {b},\vec{c}} \right)}
\right\rangle _{\psi _3^ + } = \left( {a_x + ia_y } \right)\left(
{b_x + ib_y } \right)\left( {c_x - ic_y } \right) + \left( {a_x -
ia_y } \right)\left( {b_x - ib_y } \right)\left( {c_x + ic_y }
\right)\nonumber
\\
&&\left\langle {E\left( {\vec {a},\vec {b},\vec{c}} \right)}
\right\rangle _{\psi _3^ - } = - \left( {a_x + ia_y } \right)\left(
{b_x + ib_y } \right)\left( {c_x - ic_y } \right) - \left( {a_x -
ia_y } \right)\left( {b_x - ib_y } \right)\left( {c_x + ic_y }
\right)
\end{eqnarray}
\noindent Note their coefficients are 1/2.

For the product states $\left| {000} \right\rangle ,\left| {001}
\right\rangle ,\left| {010} \right\rangle ,\left| {011}
\right\rangle ,\left| {100} \right\rangle ,\left| {101}
\right\rangle ,\left| {110} \right\rangle ,\left| {111}
\right\rangle $, the expected values are:
\begin{equation}
a_z b_z c_z, \quad  - a_z b_z c_z, \quad  - a_z b_z c_z, \quad a_z
b_z c_z, \quad  - a_z b_z c_z, \quad a_z b_z c_z, \quad a_z b_z c_z,
\quad - a_z b_z c_z
\end{equation}
\noindent respectively. If Eve takes general Bell inequality
measurements on the three uncorrelated particles, she will get 0
for a large number of measurements when the particles are randomly
distributed among the eight GHZ-basis states. If Eve does take
three correlated particles, she will also get 0 when eight
GHZ-basis states are taken with the equal probability. So Eve gets
nothing about the control key except for guessing it randomly.
Because the control key can be repeatedly used, the probability
that Eve guesses the right control key is $\left( {\frac{1}{2}}
\right)^{3N_k }$, where $3N_k $ is the number of bits in the
control key. When $N_k = 100$, the probability is $\left(
{\frac{1}{8}} \right)^{100}$, which is practically zero.
Naturally, GCORE protocol is suitable to N-qubit setting scenario,
too. $N$-qubit maximally entangled basis states are defined as
following [20]:
\begin{equation}
\left| {\psi _j^\pm } \right\rangle = \frac{1}{\sqrt 2 }\left(
{\left| j \right\rangle \left| 0 \right\rangle \pm \left| {2^{N - 1}
- j - 1} \right\rangle \left| 1 \right\rangle } \right)
\end{equation}
\noindent where $j = j_1 j_2 \cdots j_{N - 1} $ denotes binary
notations. Then there are $2^N$ different control keys, $2^N$
operations corresponding to $E_0 ,E_1 , \cdots E_{2^{N - 1}} $,
and we need $N$ quantum channels with $\frac{2^N}{2} + 1 = 2^{N -
1} + 1$ switches each. The eavesdropper Eve only guesses the right
$N$-GHZ-basis states with probability $\frac{1}{2^N}$, as the
density operation is $\rho _{AB \cdots N} =
\frac{1}{2^N}I_{2^N\times2^ N} $.

\section{GCORE using $N$-qutrit maximally entangled basis states}\label{sec2}
One of the motivations of considering a high dimensional system for
QKD is to increase the information per particle. Another context
where using higher-dimension space might be advantageous is the key
growing. However, the practical limitations might be more severe in
realistic high-dimension cryptosystems, in particular the influence
of the detector's quantum efficiency and dark count rate [21,22].
This has been discussed in the related Ref.[23]. Here, we start to
consider the qutrit quantum system.
\subsection{GCORE protocol using 2-qutrit general Bell-basis states}
Here, let's consider the simplest scenario, two particles, each
particle has three levels, i.e. a 2-qutrit system. On the whole,
concrete four processes are similar to analysis in Sec.II.B. The
recapitulation is presented in the following.  As we know, the
general Bell-basis states can be written as [24]:
\begin{equation}
 \left| {\psi _{nm} } \right\rangle = \sum\limits_j {e^{2\pi ij
/ 3}\left| j \right\rangle \otimes \left| {j + m\bmod 3}
\right\rangle / \sqrt 3 }
\end{equation}
\noindent where $n,m,j = 0,1,2$, the explicit expressions are then
\begin{eqnarray}
 &&\left| {\psi _{00} } \right\rangle = \left( {\left| {00} \right\rangle +
\left| {11} \right\rangle + \left| {22} \right\rangle } \right) / \sqrt 3 \nonumber\\
 &&\left| {\psi _{10} } \right\rangle = \left( {\left| {00} \right\rangle +
e^{2i\pi / 3}\left| {11} \right\rangle + e^{4i\pi / 3}\left| {22}
\right\rangle } \right) / \sqrt 3 \nonumber\\
 &&\left| {\psi _{20} } \right\rangle = \left( {\left| {00} \right\rangle +
e^{4i\pi / 3}\left| {11} \right\rangle + e^{2i\pi / 3}\left| {22}
\right\rangle } \right) / \sqrt 3 \nonumber\\
&&\left| {\psi _{01} } \right\rangle = \left( {\left| {01}
\right\rangle +
\left| {12} \right\rangle + \left| {20} \right\rangle } \right) / \sqrt 3 \nonumber\\
 &&\left| {\psi _{11} } \right\rangle = \left( {\left| {01} \right\rangle +
e^{2i\pi / 3}\left| {12} \right\rangle + e^{4i\pi / 3}\left| {20}
\right\rangle } \right) / \sqrt 3 \nonumber\\
 &&\left| {\psi _{21} } \right\rangle = \left( {\left| {01} \right\rangle +
e^{4i\pi / 3}\left| {12} \right\rangle + e^{2i\pi / 3}\left| {20}
\right\rangle } \right) / \sqrt 3 \nonumber\\
  &&\left| {\psi _{02} } \right\rangle = \left( {\left| {02} \right\rangle +
\left| {10} \right\rangle + \left| {21} \right\rangle } \right) / \sqrt 3 \nonumber\\
 &&\left| {\psi _{12} } \right\rangle = \left( {\left| {02} \right\rangle +
e^{2i\pi / 3}\left| {10} \right\rangle + e^{4i\pi / 3}\left| {21}
\right\rangle } \right) / \sqrt 3 \nonumber\\
 &&\left| {\psi _{22} } \right\rangle = \left( {\left| {02} \right\rangle +
e^{4i\pi / 3}\left| {10} \right\rangle + e^{2i\pi / 3}\left| {21}
\right\rangle } \right) / \sqrt 3
\end{eqnarray}
It is clear that these states are orthogonal. They can be
presented by 00, 01, 02, 10, 11, 12, 20, 21, 22, respectively. It
can be shown that single-body operators $U_{ij} \,(i,j = 0,1,2)$
will transform $\left| {\psi _{00} } \right\rangle $ into the
corresponding other eight states. The expressions of these
operators are:
\begin{eqnarray}
 U_{00} = \left( \begin{array}{*{20}c}
 1  & 0  & 0  \\
 0  & 1  & 0  \\
 0  & 0  & 1  \\
\end{array}  \right);U_{10} = \left( \begin{array}{*{20}c}
 1  & 0  & 0  \\
 0  & {e^{2\pi i / 3}}  & 0  \\
 0  & 0  & {e^{4\pi i / 3}}  \\
\end{array}  \right);U_{20} = \left( \begin{array}{*{20}c}
 1  & 0  & 0  \\
 0  & {e^{4\pi i / 3}}  & 0  \\
 0  & 0  & {e^{2\pi i / 3}}  \\
\end{array}  \right) \nonumber\\
 U_{01} = \left( \begin{array}{*{20}c}
 0  & 0  & 1  \\
 1  & 0  & 0  \\
 0  & 1  & 0  \\
\end{array} \right);U_{11} = \left( \begin{array}{*{20}c}
 0  & 0  & {e^{4\pi i / 3}}  \\
 1  & 0  & 0  \\
 0  & {e^{2\pi i / 3}}  & 0  \\
\end{array}  \right);U_{21} = \left( \begin{array}{*{20}c}
 0  & 0  & {e^{2\pi i / 3}}  \\
 1  & 0  & 0  \\
 0  & {e^{4\pi i / 3}}  & 0  \\
\end{array}  \right) \nonumber\\
 U_{02} = \left( \begin{array}{*{20}c}
 0  & 1  & 0  \\
 0  & 0  & 1  \\
 1  & 0  & 0  \\
\end{array}  \right);U_{12} = \left( \begin{array}{*{20}c}
 0  & {e^{2\pi i / 3}}  & 0  \\
 0  & 0  & {e^{4\pi i / 3}}  \\
 1  & 0  & 0  \\
\end{array}  \right);U_{22} = \left( \begin{array}{*{20}c}
 0  & {e^{4\pi i / 3}}  & 0  \\
 0  & 0  & {e^{2\pi i / 3}}  \\
 1  & 0  & 0  \\
\end{array}  \right)
\end{eqnarray}
The GCORE operations using qutrit states are similar to the cases
in Sec.II. However, there are nine choices of GCORE operations,
corresponding relations are the following
\[
\begin{array}{l}
 E_0 \leftrightarrow 00,\quad E_1 \leftrightarrow 01,\quad E_2
\leftrightarrow 02 \\
 E_3 \leftrightarrow 10,\quad E_4 \leftrightarrow 11,\quad E_5
\leftrightarrow 12 \\
 E_6 \leftrightarrow 20,\quad E_7 \leftrightarrow 21,\quad E_8
\leftrightarrow 22 \\
 \end{array}
\]
\noindent and the GCORE is done for every nine general Bell-basis
states. These operations are denoted by the denotation of
permutation group.
\[
\begin{array}{l}
 E_0 = = \left( \begin{array}{*{20}c}
 1  & 2  & 3  & 4  & 5  & 6  & 7 & 8  & 9  \\
 1  & 2  & 3  & 4  & 5  & 6  & 7 & 8  & 9  \\
\end{array}  \right) \\
~~~~ = \left( \begin{array}{*{20}c}
 1  & 1  \\
\end{array}  \right)\left( \begin{array}{*{20}c}
 2  & 2  \\
\end{array}  \right)\left( \begin{array}{*{20}c}
 3  & 3  \\
\end{array}  \right)\left( \begin{array}{*{20}c}
 4  & 4  \\
\end{array}  \right)\left( \begin{array}{*{20}c}
 5  & 5  \\
\end{array}  \right)\left( \begin{array}{*{20}c}
 6  & 6  \\
\end{array}  \right)\left( \begin{array}{*{20}c}
 7  & 7  \\
\end{array}  \right)\left( \begin{array}{*{20}c}
 8  & 8  \\
\end{array}  \right)\left( \begin{array}{*{20}c}
 9  & 9  \\
\end{array}  \right) \\
 \end{array}
\]
\[
\begin{array}{l}
 E_1 = = \left( \begin{array}{*{20}c}
 1  & 2  & 3  & 4  & 5  & 6  & 7 & 8  & 9  \\
 2  & 3  & 4  & 5  & 6  & 7  & 8 & 9  & 2  \\
\end{array}  \right) \\
 ~~~~ = \left( \begin{array}{*{20}c}
 1  & 2  \\
\end{array}  \right)\left( \begin{array}{*{20}c}
 2  & 3  \\
\end{array}  \right)\left( \begin{array}{*{20}c}
 3  & 4  \\
\end{array}  \right)\left( \begin{array}{*{20}c}
 4  & 5  \\
\end{array}  \right)\left( \begin{array}{*{20}c}
 5  & 6  \\
\end{array}  \right)\left( \begin{array}{*{20}c}
 6  & 7  \\
\end{array}  \right)\left( \begin{array}{*{20}c}
 7  & 8  \\
\end{array}  \right)\left( \begin{array}{*{20}c}
 8  & 9  \\
\end{array}  \right)\left( \begin{array}{*{20}c}
 9  & 1  \\
\end{array}  \right) \\
 \end{array}
\]
\[
\begin{array}{l}
 E_2 = = \left( \begin{array}{*{20}c}
 1  & 2  & 3  & 4  & 5  & 6  & 7 & 8  & 9  \\
 3  & 4  & 5  & 6  & 7  & 8  & 9 & 1  & 2  \\
\end{array}  \right) \\
~~~~ = \left( \begin{array}{*{20}c}
 1  & 3  \\
\end{array}  \right)\left( \begin{array}{*{20}c}
 2  & 4  \\
\end{array}  \right)\left( \begin{array}{*{20}c}
 3  & 5  \\
\end{array}  \right)\left( \begin{array}{*{20}c}
 4  & 6  \\
\end{array}  \right)\left( \begin{array}{*{20}c}
 5  & 7  \\
\end{array}  \right)\left( \begin{array}{*{20}c}
 6  & 8  \\
\end{array}  \right)\left( \begin{array}{*{20}c}
 7  & 9  \\
\end{array}  \right)\left( \begin{array}{*{20}c}
 8  & 1  \\
\end{array}  \right)\left( \begin{array}{*{20}c}
 9  & 2  \\
\end{array}  \right) \\
 \end{array}
\]
\[
\begin{array}{l}
 E_3 = = \left( \begin{array}{*{20}c}
 1  & 2  & 3  & 4  & 5  & 6  & 7 & 8  & 9  \\
 4  & 5  & 6  & 7  & 8  & 9  & 1 & 2  & 3  \\
\end{array}  \right) \\
~~~~ = \left( \begin{array}{*{20}c}
 1  & 4  \\
\end{array}  \right)\left( \begin{array}{*{20}c}
 2  & 5  \\
\end{array}  \right)\left( \begin{array}{*{20}c}
 3  & 6  \\
\end{array}  \right)\left( \begin{array}{*{20}c}
 4  & 7  \\
\end{array}  \right)\left( \begin{array}{*{20}c}
 5  & 8  \\
\end{array}  \right)\left( \begin{array}{*{20}c}
 6  & 9  \\
\end{array}  \right)\left( \begin{array}{*{20}c}
 7  & 1  \\
\end{array}  \right)\left( \begin{array}{*{20}c}
 8  & 2  \\
\end{array}  \right)\left( \begin{array}{*{20}c}
 9  & 3  \\
\end{array}  \right) \\
 \end{array}
\]
\[
\begin{array}{l}
 E_4 = = \left( \begin{array}{*{20}c}
 1  & 2  & 3  & 4  & 5  & 6  & 7 & 8  & 9  \\
 5  & 6  & 7  & 8  & 9  & 1  & 2 & 3  & 4  \\
\end{array}  \right) \\
~~~~ = \left( \begin{array}{*{20}c}
 1  & 5  \\
\end{array}  \right)\left( \begin{array}{*{20}c}
 2  & 6  \\
\end{array}  \right)\left( \begin{array}{*{20}c}
 3  & 7  \\
\end{array}  \right)\left( \begin{array}{*{20}c}
 4  & 8  \\
\end{array}  \right)\left( \begin{array}{*{20}c}
 5  & 9  \\
\end{array}  \right)\left( \begin{array}{*{20}c}
 6  & 1  \\
\end{array}  \right)\left( \begin{array}{*{20}c}
 7  & 2  \\
\end{array}  \right)\left( \begin{array}{*{20}c}
 8  & 3  \\
\end{array}  \right)\left( \begin{array}{*{20}c}
 9  & 4  \\
\end{array}  \right) \\
 \end{array}
\]
\[
\begin{array}{l}
 E_5 = = \left( \begin{array}{*{20}c}
 1  & 2  & 3  & 4  & 5  & 6  & 7 & 8  & 9  \\
 6  & 7  & 8  & 9  & 1  & 2  & 3 & 4  & 5  \\
\end{array}  \right) \\
~~~~ = \left( \begin{array}{*{20}c}
 1  & 6  \\
\end{array}  \right)\left( \begin{array}{*{20}c}
 2  & 7  \\
\end{array}  \right)\left( \begin{array}{*{20}c}
 3  & 8  \\
\end{array}  \right)\left( \begin{array}{*{20}c}
 4  & 9  \\
\end{array}  \right)\left( \begin{array}{*{20}c}
 5  & 1  \\
\end{array}  \right)\left( \begin{array}{*{20}c}
 6  & 2  \\
\end{array}  \right)\left( \begin{array}{*{20}c}
 7  & 3  \\
\end{array}  \right)\left( \begin{array}{*{20}c}
 8  & 4  \\
\end{array}  \right)\left( \begin{array}{*{20}c}
 9  & 5  \\
\end{array}  \right) \\
 \end{array}
\]
\[
\begin{array}{l}
 E_6 = = \left( \begin{array}{*{20}c}
 1  & 2  & 3  & 4  & 5  & 6  & 7 & 8  & 9  \\
 7  & 8  & 9  & 1  & 2  & 3  & 4 & 5  & 6  \\
\end{array}  \right) \\
~~~~ = \left( \begin{array}{*{20}c}
 1  & 7  \\
\end{array}  \right)\left( \begin{array}{*{20}c}
 2  & 8  \\
\end{array}  \right)\left( \begin{array}{*{20}c}
 3  & 9  \\
\end{array}  \right)\left( \begin{array}{*{20}c}
 4  & 1  \\
\end{array}  \right)\left( \begin{array}{*{20}c}
 5  & 2  \\
\end{array}  \right)\left( \begin{array}{*{20}c}
 6  & 3  \\
\end{array}  \right)\left( \begin{array}{*{20}c}
 7  & 4  \\
\end{array}  \right)\left( \begin{array}{*{20}c}
 8  & 5  \\
\end{array}  \right)\left( \begin{array}{*{20}c}
 9  & 6  \\
\end{array}  \right) \\
 \end{array}
\]
\[
\begin{array}{l}
 E_7 = = \left( \begin{array}{*{20}c}
 1  & 2  & 3  & 4  & 5  & 6  & 7 & 8  & 9  \\
 8  & 9  & 1  & 2  & 3  & 4  & 5 & 6  & 7 \ \\
\end{array}  \right) \\
~~~~ = \left( \begin{array}{*{20}c}
 1  & 8  \\
\end{array}  \right)\left( \begin{array}{*{20}c}
 2  & 9  \\
\end{array}  \right)\left( \begin{array}{*{20}c}
 3  & 1  \\
\end{array}  \right)\left( \begin{array}{*{20}c}
 4  & 2  \\
\end{array}  \right)\left( \begin{array}{*{20}c}
 5  & 3  \\
\end{array}  \right)\left( \begin{array}{*{20}c}
 6  & 4  \\
\end{array}  \right)\left( \begin{array}{*{20}c}
 7  & 5  \\
\end{array}  \right)\left( \begin{array}{*{20}c}
 8  & 6  \\
\end{array}  \right)\left( \begin{array}{*{20}c}
 9  & 7  \\
\end{array}  \right) \\
 \end{array}
\]
\[
\begin{array}{l}
 E_8 = = \left( \begin{array}{*{20}c}
 1  & 2  & 3  & 4  & 5  & 6  & 7
& 8  & 9  \\
 9  & 1  & 2  & 3  & 4  & 5 & 6 & 7  & 8  \\
\end{array}  \right) \\
 ~~~~= \left( \begin{array}{*{20}c}
 1  & 9  \\
\end{array}  \right)\left( \begin{array}{*{20}c}
 2  & 1  \\
\end{array}  \right)\left( \begin{array}{*{20}c}
 3  & 2  \\
\end{array}  \right)\left( \begin{array}{*{20}c}
 4  & 3  \\
\end{array}  \right)\left( \begin{array}{*{20}c}
 5  & 4  \\
\end{array}  \right)\left( \begin{array}{*{20}c}
 6  & 5  \\
\end{array}  \right)\left( \begin{array}{*{20}c}
 7  & 6  \\
\end{array}  \right)\left( \begin{array}{*{20}c}
 8  & 7  \\
\end{array}  \right)\left( \begin{array}{*{20}c}
 9  & 8  \\
\end{array}  \right) \\
 \end{array}
\]

\noindent the permutation has been shown clearly in Fig. 3.\quad
Fig. 4 gives the main device to perform GCORE operation by a
specific instance.

\begin{figure}[htbp]
\centering
\includegraphics[width=4.80in,height=2.80in]{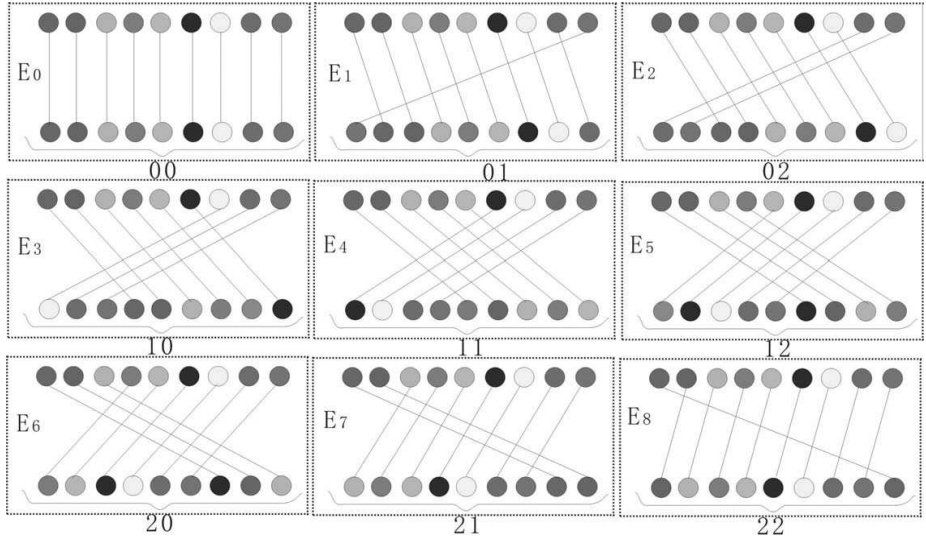}\caption{Example
of GCORE using 2-qutrit Bell-basis states;~~There are nine
different GCORE operations }
\end{figure}
\begin{figure}[htbp]
\centering
\includegraphics[width=5.30in,height=1.10in]{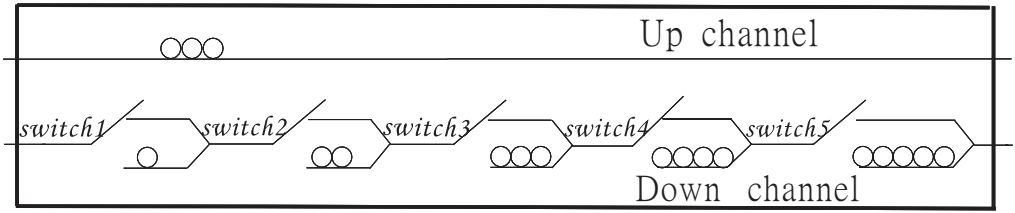}\caption{Devices
to perform GCORE operations, the loop represents a time delay of a
fixed interval.}
\end{figure}

According to Fig. 4, the upper QIC parts are transmitted according
to their temporal order. A control key is used to rearrange the
order of the lower QIC parts. For instance, the value of control
key is 00, the operation $E_0 $ is applied. In Fig. 4 there are
five switches, the order of nine general Bell-basis states is
unchanged with switch 1,2,3,4 and 5 in position (up, up, down, up,
up ). When control key is 01, $E_1 $ is performed, and it is done
by putting the nine switches into position (down, down, up, up,
down ), (up, down, up, up, up ), (up, down, up, up, up ), (up,
down, up, up, up ), (up, down, up, up, up ), (up, down, up, up, up
), (up, down, up, up, up ), (up, down, up, up, up ), (up, down,
up, up, up ) for the nine particles, respectively. Similar
combination can be written explicitly for $E_2, E_3, E_4, E_5,
E_6, E_7, E_8$.
Now we can consider the cases of multi-particle
and/or high-dimension quantum systems. Firstly the method is
generalized to high dimension quantum system $\left( {d
> 3} \right)$ of two particles. Note that the $d$-dimension Bell-basis states in a symmetric
channel [8, 21, 23] are expressed as
\begin{equation}
\left| {\psi _{nm} } \right\rangle = \sum\limits_j {e^{2\pi ijn /
d}\left| j \right\rangle \otimes \left| {j + m\bmod d} \right\rangle
/ \sqrt d }
\end{equation}
\noindent where $n,m,j = 0,1, \cdots d - 1$. The unitary operator
is
\begin{equation}
U_{nm} \, = \sum\limits_j {e^{2\pi ijn / d}\left| {j + m\bmod d}
\right\rangle \left\langle j \right|}
\end{equation}
\noindent which can transfer $d$-dimension Bell-basis state
\begin{equation}
\left| {\psi _{00} } \right\rangle = \sum\limits_j {\left| j
\right\rangle \otimes \left| j \right\rangle / \sqrt d }
\end{equation}
\noindent to other $d$-dimension Bell-basis state $\left| {\psi
_{nm} } \right\rangle $, i.e. $U_{nm} \left| {\psi _{00} }
\right\rangle \, = \left| {\psi _{nm} } \right\rangle $. So we can
use the same method like 2-qutrit GCORE to analyze this problem
completely.
Thus, we have presented the GCORE of two-particle high
dimensional generalization, we will give multi-particle situation
next. At first, we consider a less complicated three particle
quantum system. For 3-qutrit quantum system, its generalized
maximally entangled basis states are :
\begin{equation}
\left| {\psi _{nm}^k } \right\rangle = \sum\limits_j {e^{2\pi ijk /
3}\left| j \right\rangle \otimes \left| {j + n\bmod 3} \right\rangle
\otimes \left| {j + m\bmod 3} \right\rangle / \sqrt 3 }
\end{equation}
\noindent where $n,m,k = 0,1,2$, the explicit expressions are then
\begin{eqnarray}
 &&\left| {\psi _{00}^0 } \right\rangle = \left( {\left| {000} \right\rangle +
\left| {111} \right\rangle + \left| {222} \right\rangle } \right) /
\sqrt 3\nonumber
\\
 &&\left| {\psi _{01}^0 } \right\rangle = \left( {\left| {001} \right\rangle +
\left| {112} \right\rangle + \left| {220} \right\rangle } \right) /
\sqrt 3\nonumber
\\
 &&\left| {\psi _{02}^0 } \right\rangle = \left( {\left| {002} \right\rangle +
\left| {110} \right\rangle + \left| {221} \right\rangle } \right) /
\sqrt 3\nonumber
\\
 && \cdots \nonumber\\
 &&\left| {\psi _{22}^2 } \right\rangle = \left( {\left| {022} \right\rangle +
e^{4i\pi / 3}\left| {100} \right\rangle + e^{2i\pi / 3}\left| {222}
\right\rangle } \right) / \sqrt 3
\end{eqnarray}
\noindent  There are 27 corresponding GCORE operations, denoted
by:
\begin{eqnarray}
E_0 \leftrightarrow 000,~~E_1 \leftrightarrow 001,~~E_2
\leftrightarrow 002,~~E_3 \leftrightarrow 100,~~E_4 \leftrightarrow
101,~~E_5\leftrightarrow 102,~~E_6 \leftrightarrow 200, \nonumber\\
E_7 \leftrightarrow 201,~~E_8 \leftrightarrow 202,~~E_9
\leftrightarrow 010,~~E_{11} \leftrightarrow 011,~~E_{11}
\leftrightarrow 012,~~E_{12}
\leftrightarrow 110,~~E_{13} \leftrightarrow 111, \nonumber\\
E_{14} \leftrightarrow 112,~~E_{15} \leftrightarrow 210,~~E_{16}
\leftrightarrow 211,~~E_{17} \leftrightarrow 212,~~E_{18}
\leftrightarrow
020,~~E_{19} \leftrightarrow 021,~~E_{20} \leftrightarrow 022, \nonumber\\
\noindent E_{21} \leftrightarrow 120,~~E_{22} \leftrightarrow
121,~~E_{23} \leftrightarrow 122,~~E_{24} \leftrightarrow
220,~~E_{25}
\leftrightarrow 221,~~E_{26} \leftrightarrow 222 \nonumber\\
\end{eqnarray}
Due to the complication of GCORE operations, more resources are
needed, and the analysis of security also becomes more
complicated. But the maximal advantage is the swell of security.
And the probability that Eve guesses the right control key is near
0. The corresponding fig. 5 is given below.
\begin{figure}[h]
\centering
\includegraphics[width=5.30in,height=3.04in]{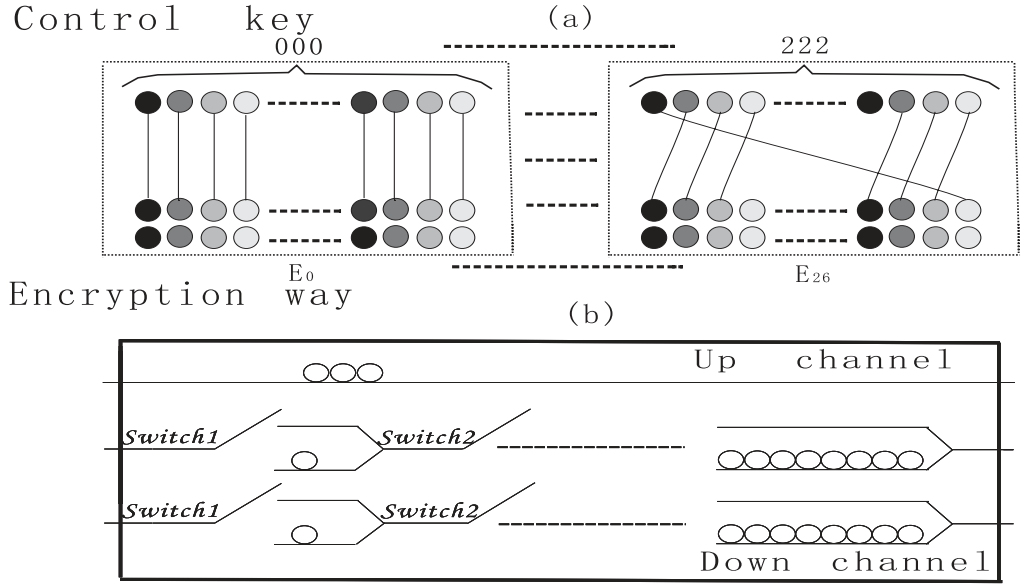}\caption{(a).
Example of GCORE using 3-qutrit maximally entangled basis states;
(b). Devices to perform GCORE operations, the loop represents a
time delay of a fixed interval.}
\end{figure}

Generally, a uniform expression of $N$-qutrit maximally entangled
basis state can be expressed as the following form
\begin{eqnarray}
 |{\psi _{i_1 ,i_2 , \cdots ,i_{N-1} }^N }\rangle &=&\nonumber
\sum\limits_je^{2\pi ijN / 3}| j \rangle \otimes| {j + i_1 \bmod
3}\rangle \otimes| {j + i_2 \bmod 3}\rangle\\
& &\otimes \cdots \otimes | {j + i_{N-1} \bmod 3}\rangle / \sqrt{3}
\end{eqnarray}
\noindent where $i_1,i_2\cdots i_N= 0,1,2$. Similar analysis can
be given, but there is a little difference. In short, there are
$3^N$ different control keys, $3^N$ operations corresponding to
$E_0 ,E_1 , \cdots E_{3^N- 1}$, and we need $N$ quantum channels
with $3^{N - 1} + 1$ switches each. The eavesdropper only guesses
the right general Bell-basis state with probability
$\frac{1}{3^N}$, as the density operation is $\rho _{AB \cdots N}
= \frac{1}{3^N}I_{3^N\times 3^N} $.

\subsection{Security of GCORE using 2-qutrit general Bell-basis states}
Now, let us look at the security of above GCORE protocol using
2-qutrit states. Eve has only $11.1\%$ chance to guess the right
GCORE operation for nine general Bell-basis states. If she uses a
wrong GCORE operation, the two particles she measured will be
anticorrelated. Assume that A particle from the general Bell-basis
state, B particle from the second general Bell-basis state are
mistreated by Eve as a general Bell-basis state, then the density
operator will be
\begin{equation}
\rho _{A_1B_2} = \bar {\rho }_{A_1 } \otimes \bar {\rho }_{B_2 } =
\left( \begin{array}{*{20}c}
 {\displaystyle\frac{1}{3}}  & 0  & 0  \\
 0  & {\displaystyle\frac{1}{3}}  & 0  \\
 0  & 0  & {\displaystyle\frac{1}{3}}  \\
\end{array}  \right) \otimes \left( \begin{array}{*{20}c}
 {\displaystyle\frac{1}{3}}  & 0  & 0  \\
 0  & {\displaystyle\frac{1}{3}}  & 0  \\
 0  & 0  & {\displaystyle\frac{1}{3}}  \\
\end{array}  \right) = \frac{1}{9}I_{9\times 9}
\end{equation}
\noindent where $\bar {\rho }_{A_1 } = \Tr_{B_1 } \left( {\rho
_{A_1 B_1 } } \right),\bar {\rho }_{B_2 } = \Tr_{A_2 } \left(
{\rho _{A_2 B_2 } } \right).$

The result indicates that any one of
the nine general Bell-basis states appears with $11.1\%$
probability each. Thus Eve will introduce $79.01\% $ error rates
in the results. Alice and Bob can detect Eve easily by checking a
sufficiently large subset of results randomly chosen. Surely, Eve
can take the generalized Bell inequality measurement on the
particles, but it is useless for decrypting the control key.
There are eight (Hermitian) generators of $SU\left( 3 \right)$, i.e.
eight Gell-Mann matrices, which are defined by
\[
\left( \begin{array}{*{20}c}
 0  & 1  & 0  \\
 1  & 0  & 1  \\
 0  & 1  & 0  \\
\end{array}  \right),
\left( \begin{array}{*{20}c}
 0  & { - i}  & 0  \\
 i  & 0  & 0  \\
 0  & 0  & 0  \\
\end{array}  \right),
\left( \begin{array}{*{20}c}
 1  & 0  & 0  \\
 0  & { - 1}  & 0  \\
 0  & 0  & 0  \\
\end{array}  \right),
\left( \begin{array}{*{20}c}
 0  & 0  & 1  \\
 0  & 0  & 0  \\
 1  & 0  & 0  \\
 \end{array}  \right)
\]
\[
\left( \begin{array}{*{20}c}
 0  & 0  & { - i}  \\
 0  & 0  & 0  \\
 i  & 0  & 0  \\
\end{array}  \right),
\left( \begin{array}{*{20}c}
 0  & 0  & 0  \\
 0  & 0  & 1  \\
 0  & 1  & 0  \\
\end{array}  \right) \\,
\left( \begin{array}{*{20}c}
 0  & 0  & 0  \\
 0  & 0  & { - i} \\
 0  & i  & 0  \\
\end{array}  \right),
\displaystyle\frac{1 }{\sqrt 3 }\left( \begin{array}{*{20}c}
 1  & 0  & 0  \\
 0  & 1  & 0  \\
 0  & 0  & { - 2}  \\
\end{array}  \right) \\
\]
Let us choose directions $\vec {M}$ and $\vec {N}$ as the
directions of measurement of Alice and Bob respectively, these
measurements satisfy the orthogonal relations. The correlation
operator can be written as:
\begin{equation}
\hat {E} = \hat {S} \cdot \vec {M} \otimes \hat {S} \cdot \vec {N}
\end{equation}

The expectation values $\left\langle {E\left( {\vec {M},\vec {N}}
\right)} \right\rangle _\psi = \left\langle \psi \right|\hat {S}
\cdot \vec {M} \otimes \hat {S} \cdot \vec {N}\left| \psi
\right\rangle $ are not equal for different general Bell-basis
states.

\begin{eqnarray}
\left\langle {E\left( {\vec {M},\vec {N}} \right)} \right\rangle
_{\psi _{00} } = \quad \frac{2}{3}\left( {\begin{array}{l}
 M_1 N_1 - M_2 N_2 + M_3 N_3 + M_4 N_4 - M_5 N_5 \\
 + M_6 N_6 - M_7 N_7 + M_8 N_8 \\
 \end{array}} \right)\nonumber
\\
\left\langle {E\left( {\vec {M},\vec {N}} \right)} \right\rangle
_{\psi _{01} } = \frac{1}{3}\left( {\begin{array}{l}
 2M_4 N_1 + 2M_5 N_2 - M_3 N_3 - \sqrt 3 M_8 N_3 + 2M_6 N_4 \\
 + 2M_7 N_5 + 2M_1 N_6 - 2M_2 N_7 + \sqrt 3 M_3 N_8 - M_8 N_8 \\
 \end{array}} \right)\nonumber
\\
\left\langle {E\left( {\vec {M},\vec {N}} \right)} \right\rangle
_{\psi _{02} } = \frac{1}{3}\left( {\begin{array}{l}
 2M_6 N_1 - 2M_7 N_2 - M_3 N_3 + \sqrt 3 M_8 N_3 + 2M_1 N_4 \\
 + 2M_2 N_5 + 2M_4 N_6 + 2M_5 N_7 - \sqrt 3 M_3 N_8 - M_8 N_8 \\
 \end{array}} \right)\nonumber
\\
\begin{array}{l} \left\langle {E\left( {\vec {M},\vec {N}}
\right)} \right\rangle _{\psi _{10} } = \frac{ - 1 + i\sqrt 3
}{3}M_1 N_1 - \frac{ - 1 + i\sqrt 3 }{3}M_2
N_2 + \frac{1 - i\sqrt 3 }{6}M_3 N_3 + \frac{\sqrt 3 + i}{6}M_8 N_3 \\
 \quad \quad \quad {\kern 1pt} \quad \quad \quad\quad\quad - \frac{1 + i\sqrt 3
}{3}M_4 N_4 + \frac{1 + i\sqrt 3 }{3}M_5 N_5 + \frac{2}{3}M_6 N_6 -
\frac{2}{3}M_7 N_7  \\
 \quad \quad \quad \quad \quad \quad\quad\quad\quad + \frac{\sqrt 3 + i}{6}M_3 N_8- \frac{1 - i\sqrt 3 }{6}M_8 N_8 \nonumber\\
 \end{array}
 \\
\begin{array}{l}
 \left\langle {E\left( {\vec {M},\vec {N}} \right)} \right\rangle _{\psi
_{11} } = - \frac{1 + i\sqrt 3 }{3}M_4 N_1 - \frac{1 + i\sqrt 3
}{3}M_5 N_2
- \frac{1}{3}M_3 N_3 - \frac{i}{3}M_8 N_3 \\
 \quad \quad \quad {\kern 1pt} \quad \quad \quad\quad\quad + \frac{2}{3}M_6 N_4 +
\frac{2}{3}M_7 N_5 + \frac{ - 1 + i\sqrt 3 }{3}M_1 N_6 + \frac{1 -
i\sqrt 3
}{3}M_2 N_7 \\
 \quad \quad \quad \quad \quad \quad\quad\quad - \frac{i}{3}M_3 N_8 + \frac{1}{3}M_8
N_8 \nonumber\\
 \end{array}
\\
\begin{array}{l}
 \left\langle {E\left( {\vec {M},\vec {N}} \right)} \right\rangle _{\psi
_{12} } = \frac{2}{3}M_6 N_1 - \frac{2}{3}M_7 N_2 + \frac{1 + i\sqrt
3
}{6}M_3 N_3 - \frac{\sqrt 3 - i}{6}M_8 N_3 \\
 \quad \quad \quad {\kern 1pt} \quad \quad \quad\quad\quad - \frac{1 - i\sqrt 3
}{3}M_1 N_4 - \frac{1 - i\sqrt 3 }{3}M_2 N_5 - \frac{1 + i\sqrt 3
}{3}M_4
N_6 \\
 \quad \quad \quad \quad \quad \quad\quad\quad - \frac{1 + i\sqrt 3 }{3}M_5 N_7 -
\frac{\sqrt 3 - i}{6}M_3 N_8 - \frac{1 + i\sqrt 3 }{6}M_8 N_8 \nonumber\\
 \end{array}
 \\
\begin{array}{l}
 \left\langle {E\left( {\vec {M},\vec {N}} \right)} \right\rangle _{\psi
_{20} } = - \frac{1 + i\sqrt 3 }{3}M_1 N_1 + \frac{1 + i\sqrt 3
}{3}M_2 N_2
+ \frac{1 + i\sqrt 3 }{6}M_3 N_3 + \frac{\sqrt 3 - i}{6}M_8 N_3 \\
 \quad \quad \quad {\kern 1pt} \quad \quad \quad\quad\quad - \frac{1 - i\sqrt 3
}{3}M_4 N_4 + \frac{1 - i\sqrt 3 }{3}M_5 N_5 + \frac{2}{3}M_6 N_6 -
\frac{2}{3}M_7 N_7 \\
 \quad \quad \quad \quad \quad \quad\quad\quad + \frac{\sqrt 3 - i}{6}M_3 N_8 -
\frac{1 + i\sqrt 3 }{6}M_8 N_8 \nonumber\\
 \end{array}
\\
\begin{array}{l}
 \left\langle {E\left( {\vec {M},\vec {N}} \right)} \right\rangle _{\psi
_{21} } = \frac{ - 1 + i\sqrt 3 }{3}M_4 N_1 + \frac{ - 1 + i\sqrt 3
}{3}M_5
N_2 - \frac{1}{3}M_3 N_3 + \frac{i}{3}M_8 N_3 \\
 \quad \quad \quad {\kern 1pt} \quad \quad \quad\quad\quad + \frac{2}{3}M_6 N_4 +
\frac{2}{3}M_7 N_5 - \frac{1 + i\sqrt 3 }{3}M_1 N_6 + \frac{1 +
i\sqrt 3
}{3}M_2 N_7 \\
 \quad \quad \quad \quad \quad \quad\quad\quad + \frac{i}{3}M_3 N_8 + \frac{1}{3}M_8
N_8 \nonumber\\
 \end{array}
\\
\begin{array}{l}
 \left\langle {E\left( {\vec {M},\vec {N}} \right)} \right\rangle _{\psi
_{22} } = \frac{2}{3}M_6 N_1 - \frac{2}{3}M_7 N_2 + \frac{1 - i\sqrt
3
}{6}M_3 N_3 - \frac{\sqrt 3 + i}{6}M_8 N_3 \\
 \quad \quad \quad {\kern 1pt} \quad \quad \quad\quad\quad - \frac{1 + i\sqrt 3
}{3}M_1 N_4 - \frac{1 + i\sqrt 3 }{3}M_2 N_5 - \frac{1 - i\sqrt 3
}{3}M_4
N_6 \\
 \quad \quad \quad \quad \quad \quad\quad\quad - \frac{1 - i\sqrt 3 }{3}M_5 N_7 -
\frac{\sqrt 3 + i}{6}M_3 N_8 - \frac{1 - i\sqrt 3 }{6}M_8 N_8 \\
 \end{array}
\end{eqnarray}

\noindent For product states $\left| {00} \right\rangle ,\left|
{01} \right\rangle ,\left| {02} \right\rangle ,\left| {10}
\right\rangle ,\left| {11} \right\rangle ,\left| {12}
\right\rangle ,\left| {20} \right\rangle ,\left| {21}
\right\rangle ,\left| {22} \right\rangle $, the expected values
are
\begin{eqnarray*}
\left( {M_3 + \frac{M_8 }{\sqrt 3 }} \right)\left( {N_3 + \frac{N_8
}{\sqrt 3 }} \right), ~\left( {M_3 + \frac{M_8 }{\sqrt 3 }}
\right)\left( { - N_3 + \frac{N_8 }{\sqrt 3 }} \right),
 ~- \left( {M_3 + \frac{M_8 }{\sqrt 3 }} \right)\frac{N_8 }{\sqrt 3 },
 \\
\left( { - M_3 + \frac{M_8 }{\sqrt 3 }} \right)\left( {N_3 +
\frac{N_8 }{\sqrt 3 }}, ~~\right), \left( {\frac{M_8 }{\sqrt 3 } -
M_3 } \right)\left( {\frac{N_8 }{\sqrt 3 } - N_3 } \right),-
\frac{2}{\sqrt 3 }\left( { - M_3 + \frac{M_8 }{\sqrt 3 }}
\right)N_8,
\\
- \frac{2}{\sqrt 3 }\left( {N_3 + \frac{N_8 }{\sqrt 3 }} \right)N_8
,  - \frac{2}{\sqrt 3 }\left( { - N_3 + \frac{N_8 }{\sqrt 3 }}
\right)M_8 , ~~\frac{4}{3}M_8 N_8
\end{eqnarray*}

\noindent respectively. Subsequently, it's easy to give the
similar analysis like CORE using EPR pairs.

The experimental realization about qutrit for quantum cryptography
is important. Up to now, the experiment has achieved much progress
to realize the production of general Bell-basis state [25,26]. For
example, in order to produce the state
\begin{equation}\left| {\psi _{00} } \right\rangle = \left( {\left| {00}
\right\rangle + \left| {11} \right\rangle + \left| {22}
\right\rangle } \right) / \sqrt 3
\end{equation}

\noindent one uses a unbiased six-port beam splitter [27] which is
a device with the following property: if a photon enters any
single input port (out of the three ports), there is equal
probability that it leaves one of the three output ports to
produce state. In fact, one can always construct a special
six-port beam splitter with the distinguishing trait that the
elements of its unitary transition matrix, $T$, are solely powers
of the complex number, $\alpha = \exp (i\frac{2\pi }{3}),$~namely,
$T_{kl} = \frac{1}{\sqrt 3 }\alpha ^{\left( {k - 1} \right)\left(
{l - 1} \right)}$. It has been shown in Ref.[22] that any six-port
beam splitter can be constructed from the above-mentioned one by
adding appropriate phase shifters at its exit and input ports (and
by a trivial relabeling of the output ports). The phase shifters
in front of the input ports of beam splitter can be tunable and
used to change the phase of the incoming photon.

\subsection{Security analysis of qutrit GCORE using the quantum
cloning machine} From now on, we will analyze the security of
qutrit GCORE against individual attacks (where Eve monitors the
qutrit separately). So far, a lot work about the analysis of
security for BB84 or generalized BB84 protocol using cloning
machine have been done [18,19,28]. These workers are
significative. Fortunately, GCORE protocols are also propitious to
analyze using these methods. For this case, we consider a fairly
general class of eavesdropping attack based on (not necessarily
universal) quantum cloning machine. It is known that such a
cloning-based attack is the optimal eavesdropping strategy, that
is, the best Eve can do is to clone (imperfectly) Alice's qubit
and keep a copy while sending the original to Bob [18]. An
appropriate measurement of the clone (and the ancilla system)
after disclosure of the basis enables Eve to gain the maximally
possible information on Alice's key bit.

We use a general class of
cloning transformations which is defined in Refs.[18,19], the
resulting joint state of the two clones (noted A and B) and of the
cloning machine (noted C) is
\begin{eqnarray}
 &&\left| \psi \right\rangle \to \sum\limits_{m,n = 0}^{N - 1} {a_{m,n}
U_{m,n} \left| \psi \right\rangle _A \left| {B_{m, - n} }
\right\rangle
_{B,C} } \nonumber\\
 && ~~~~~~= \sum\limits_{m,n = 0}^{N - 1} {b_{m,n} U_{m,n} \left| \psi
\right\rangle _B \left| {B_{m, - n} } \right\rangle _{A,C} }
\end{eqnarray}
\noindent where
\begin{equation}
U_{m,n} = \sum\limits_{k = 0}^{N - 1} {e^{2\pi i\left( {kn / N}
\right)}\left| {k + m} \right\rangle \left\langle k \right|}
\end{equation}
\noindent $U_{m,n} $ forms a group of qudit error operators,
generalizing the Pauli matrices for qubit: $m$ labels the shift
errors (extending the bit flip $\sigma _x )$, while $n$ labels the
phase errors (extending the phase flip $\sigma _z )$. And
\begin{equation}
\left| {B_{m,n} } \right\rangle = N^{ - \frac{1}{2}}\sum\limits_{k =
0}^{N - 1} {e^{2\pi i\left( {kn / N} \right)}\left| k \right\rangle
\left| {k + m} \right\rangle }
\end{equation}
\noindent with $0 \le m,n \le N - 1$. Equation $\left| {B_{m,n} }
\right\rangle $ defines the $N^2$ generalized Bell states for a
pair of $N$-dimensional systems. The final states of clone A, B
are
\begin{eqnarray}
 &&\rho _A = \sum\limits_{m,n = 0}^{N - 1} {p_{m,n} \left| {\psi _{m,n} }
\right\rangle \left\langle {\psi _{m,n} } \right|} =
\sum\limits_{m,n = 0}^{N - 1} {p_{m,n} U_{m,n} \left| \psi
\right\rangle \left\langle \psi
\right|U_{_{m,n} }^\dag } \nonumber\\
 &&\rho _B = \sum\limits_{m,n = 0}^{N - 1} {q_{m,n} \left| {\psi _{m,n} }
\right\rangle \left\langle {\psi _{m,n} } \right|} =
\sum\limits_{m,n = 0}^{N - 1} {q_{m,n} U_{m,n} \left| \psi
\right\rangle \left\langle \psi \right|U_{_{m,n} }^\dag }
\end{eqnarray}
\noindent In addition, the weight functions of the two clones are
related by
\begin{equation}
p_{m,n} = \left| {a_{m,n} } \right|^2,\quad q_{m,n} = \left|
{b_{m,n} } \right|^2
\end{equation}
\noindent where $a_{m,n} $, $b_{m,n} $ are two (complex) amplitude
functions that are dual under a Fourier transform:
\begin{equation}
b_{m,n} = \frac{1}{N}\sum\limits_{x,y = 0}^{N - 1} {e^{2\pi i\left(
{nx - my} \right) / N}a_{m,n} }
\end{equation}
Assume that Eve clones the qutrit state that is sent to Bob. Then
Eve will measure her clone in the same basis as Bob and her
ancilla in the conjugate basis. For deriving Eve's information, we
need first to rewrite the cloning transformation of these bases.
If Alice sends any state $\left| k \right\rangle $ in the
computational basis, the phase errors clearly do not play any role
in the mixture $\rho _B $, so the fidelity can be expressed as:
\begin{equation}
F = \left\langle k \right|\rho _B \left| k \right\rangle =
\sum\limits_{n = 0}^{N - 1} {\left| {a_{0,n} } \right|^2}
\end{equation}

In the rest of this subsection, we will use this general
characterization of cloning in order to investigate the
state-dependent cloning of qutrit. Alice sends the input state
$\left| \psi \right\rangle $ belonging to a 3-dimensional space.
For the cloner to copy equally well the states of computational
bases, we choose the amplitude $a_{m,n}$ characterizing the
cloner, which must be of the form
\begin{equation}
\left( {a_{m,n} } \right) = \left( \begin{array}{*{20}c}
 v  & x  & x  \\
 y  & y  & y  \\
 z  & z  & z  \\
\end{array}  \right)
\end{equation}
\noindent such a cloner is phase covariant, which means it acts
identically on each state of the computational base.

 The fidelity
of the first clone (the one that is sent to Bob) when copying a
state $\left| \psi \right\rangle $ can be written, in general, as
\begin{equation}
F_A = \left\langle \psi \right|\rho _A \left| \psi \right\rangle =
\sum\limits_{m,n = 0}^{N - 1} {\left| {a_{m,n} } \right|^2\left|
{\left\langle \psi \right|\left. {\psi _{m,n} } \right\rangle }
\right|^2} = \sum\limits_{m,n = 0}^{N - 1} {\left| {\left\langle
\psi \right|U_{m,n} \left| {\psi _{m,n} } \right\rangle } \right|^2}
\end{equation}
\noindent That is $F_A = v^2 + y^2 + z^2$. The Disturbances
$D_{A_1 } $and $D_{A_2 } $of the first clone are:
\begin{equation}
D_{A_1 } = D_{A_2 } = x^2 + y^2 + z^2
\end{equation}
\noindent By view of equation
\begin{equation}
b_{m,n} = \frac{1}{N}\sum\limits_{x,y = 0}^{N - 1} {e^{2\pi i\left(
{nx - my} \right) / N}a_{m,n} }
\end{equation}
\noindent we can obtain that, for the second clone, which is the
maximum when $y = z$, and the fidelity is given by
\begin{equation}
F_B = \left( {v^2 + 2x^2 + 12y^2 + 8xy + 4vy} \right) / 3
\end{equation}
\noindent Again, we get the same disturbances (minimal when $y =
z)$ given by
\begin{equation}
D_{B_1 } = D_{B_2 } = \left( {v^2 + 2x^2 + 3y^2 - 4xy - 2vy} \right)
/ 3
\end{equation}
For simplicity, it's natural to consider the following amplitude
matrix [19]
\begin{equation}
\left( {a_{m,n} } \right) = \left( {{\begin{array}{*{20}c}
 v  & x  & x  \\
 x  & x  & x \\
 x  & x  & x  \\
\end{array} }} \right)
\end{equation}
\noindent where $v,x$ are real parameters that satisfy the
normalization condition $v^2 + 8x^2 = 1$. It's easy to check that
this cloner's results in the same fidelity and same disturbance
for any qutrit state:
\begin{equation}
F = v^2 + 2x^2\quad \quad D_1 = D_2 = 3x^2
\end{equation}
\noindent Of course we have the relation: $F + D_1 + D_2 = 1$. We
can easily know that the symmetric universal qutrit cloner is
characterized by a fidelity of $3/4$.
Now, it is simple to analyze
its security against an incoherent attack. Bob's fidelity is $F =
v^2 + 2x^2$ and the corresponding mutual information between Alice
and Bob (if the latter measures his clone in the good basis) [18]
is given by
\begin{equation}
I_{AB} = \log _2 3 + F\log _2 F + (1 - F)\log _2 \frac{1 - F}{2}
\end{equation}
\noindent since two possible errors are equiprobable. The cloning
fidelity for Eve is given by
\begin{equation}
F_E = \frac{\left( {v + 8x} \right)^2 + 2\left( {v - x}
\right)^2}{9}
\end{equation}
\noindent Maximizing Eve's fidelity using the normalization
relation $v^2 + 8x^2 = 1$ yields the optimal cloner
\begin{equation}
x = \sqrt {\frac{F\left( {1 - F} \right)}{2}} ,\quad \quad \quad v =
F
\end{equation}
\noindent The corresponding optimal fidelity for Eve is
\begin{equation}
F_E = \frac{F}{3} + \frac{2}{3}\left( {1 - F} \right) +
\frac{2}{3}\sqrt {2F\left( {1 - F} \right)}
\end{equation}
Let us see how Eve can maximize her information on Alice's state.
If Alice sends the state $\left| k \right\rangle \left( {k =
0,1,2} \right)$, then it is clear that Eve can obtain Bob's error
simply by performing a practical Bell measurement (measuring only
the $m$ index) on BC. In order to infer Alice's state, Eve must
distinguish between three states ($\left| 0 \right\rangle ,\left|
1 \right\rangle ,\left| 2 \right\rangle )$ with a same scalar
product $\frac{3F - 1}{2}$ for all pairs of states, regardless of
the measured value of $m$. Consequently, Eve's information [18] is
\begin{equation}
I_{AE} = \log _2 3 + F_E \log _2 F_E + (1 - F_E )\log _2 \frac{1 -
F_E }{2}
\end{equation}

As a result, Bob's and Eve's information curves intersect exactly
where the fidelities coincide. That is, at   $F = F_E =
\frac{1}{2}\left( {1 + \frac{1}{\sqrt 3 }} \right) $.
\begin{figure}[h]
\centering
\includegraphics[scale=0.9]{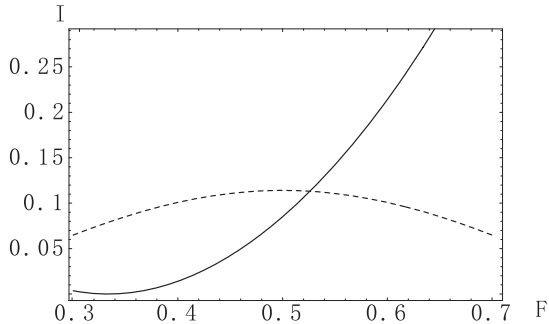}\caption{Relation
between fidelity and mutual information in which real curve
represents mutual information between Alice and Eve; the dashed
curve represents mutual information between Alice and Bob.}
\end{figure}

Due to a theorem given by Csiszar and Korner [21], which provides
a lower bound on the secret key rate. Concretely, it is sufficient
that $I_{AB} > I_{AC}$ in order to establish a secret key with a
nonzero rate, if the one-way communication on the classical
channel is used, this is actually a necessary condition.
Consequently, the GCORE protocols cease to generate secret key
bits precisely at the point where Eve's information attains Bob's
information.

We compute the disturbance $D_{qutrit} = 1 - F =
\frac{1}{2}\left( {1 - \frac{1}{\sqrt 3 }} \right) $ (or error
rate) at which $I_{AB} = I_{AE} \;(or\;F = F_E )$, that is, above
which Alice and Bob can not distill a secret key any more by use
of one--way privacy amplification protocol. While the disturbance
for the protocol using qubit is $D_{qubit} = \frac{1}{2}\left( {1
- \frac{1}{\sqrt 2 }} \right)$, we can see $D_{qubit} < D_{qutrit}
$ easily. Thus we say that disturbance increases with the
dimension, suggesting mutual information between Alice and Eve of
qutrit cryptosystem is getting smaller than that of qubit
cryptosystem under the same condition. In other words, Eve obtains
less information in the qutrit scheme. Our analysis thus confirms
a seemingly general property that qutrit scheme for QKD to be more
robust against eavesdropping than the corresponding qubit scheme.

\section {Analysis and Conclusion}\label {sec4}
For $m$ particles and/or higher dimension quantum systems, we can
provide the uniform expression of maximally entangled basis states.
Assume the number of particle is $n$, each dimension is $d$, the
maximally entangled basis states are
\begin{equation}
\left| {\psi _{i_1 ,i_2 , \cdots i_n }^n } \right\rangle =
\sum\limits_j {e^{2\pi ijn / d}\left| j \right\rangle \otimes \left|
{j + i_1 \bmod d} \right\rangle \otimes \left| {j + i_2 \bmod d}
\right\rangle \otimes \cdots \left| {j + i_n \bmod d} \right\rangle
/ \sqrt d }
\end{equation}
\noindent Similar analysis can be used, even if the dimension is
not limited to $2, 3$. If Eve wants to measure the states without
disturbing the system only if they are eigenstates of the
measuring operator, otherwise, she will produce errors at most of
time. Meantime, Eve can only guess the control key randomly, she
has no means to decipher the control key. In a word, the security
of GCORE operation becomes better than ever. Like other QKD
protocols using orthogonal states, one distinct feature of our
scheme is its high efficiency. The information-theoretic
efficiency defined in Ref.[16] is:
\begin{equation}
\eta = \frac{b_s }{q_t + b_t }
\end{equation}
\noindent where $b_s $ is the number of secret bits received by
Bob, $q_t $ is the number of qubits used, and $b_t $ is the number
of classical bits exchanged between Alice and Bob during the QKD
process. The efficiency of any protocol for QKD, defined as the
number of secret (i.e. allowing eavesdropping detection) bits per
transmitted bit plus qubit, satisfies $\eta \le 1$. The protocol
presented here becomes 100{\%}, because $b_s = d^N\log _2 d^N$,
$q_t = d^N\log _2 d^N$, $b_t = 0$. In this way, we can calculate
out that the efficiency of BB84 is 25{\%}, similarly the EPR
protocol is 50{\%}. To the best of our knowledge, only two
protocols reach the limit value of $\eta = 1$, one protocol by
Cabello (high capacity Cabello protocol, HCCP) [16] and one by
Long and Liu (high capacity Long Liu protocol HCLLP) [17]. Both
protocols exploit the fact that a possible eavesdropper with no
access to the whole quantum system at the same time, cannot
recover the whole information without being detected, and both
employ a larger alphabet, a few-dimensional orthogonal basis of
pure state. The GCORE has the same characters, so we can also
obtain the full efficiency from this point of view.

Another feature of the scheme is its high capacity since the four
possible states of the EPR pairs carry two bits of
information($\log _2 4 = 2)$, eight possible states of GHZ-basis
states carry three bits of information($\log _2 2^3 = 3)$.
Similarly, the nine possible states of the 2-qutrit general
Bell-basis states carry $\log _2 9$ bits of information, the 27
possible states of the 3-qutrit general maximally entangled basis
states carry $\log _2 3^3$ bits of information, so we can think
the possible states of the $N$-qudit maximally entangled basis
state carry $\log _2 d^N$ bits of information. In short, $M$
adopted $N$-qudit maximally entangled state can send $M\log _2
d^N$ bits of information in our GCORE scheme. On average, per
particle of GCORE protocol carries $(\log_2 d^N)/Nd^{N}$ bits of
information. Whereas in the EPR scheme (BB84) each adopted EPR
pair (particles) carried only one bit of information, that is, 0.5
bit information per particle carries. But if we use the control
key to control the GCORE operation of a group of units. We can
save a large amount of resources. From this sense, we think the
proposed scheme is better.

In QKD, our scheme is just one-to-one protocol, there are other
protocols using different ways to distribute secret keys
[1-3,6,7-17]. As we know, Townsend's protocol [30] is a one-to-any
protocol, where Alice acts as a single controller to establish and
update a distinct secret key with each network user. An any-to-any
protocol has been proposed to allow any two users to establish a
secret key over an optical network by Phoenix et al.[31]. The
present scheme can be generalized to distribute secret keys to
multiple legitimate users. It is different from Townsend and
Phoenix's protocol in that the secret keys are common to all
legitimate users. The procedure is given in the following. We
demonstrate it using EPR pair for simplicity, after Alice has sent
the keys to Bob, Bob can create an EPR pair sequence that carries
the raw keys. Then he sends this EPR pair sequence to another
legitimate user, Clare, using the same procedure and device as
before. The key protocols common to Alice, Bob and Clare are those
Bell-basis measurement results that are not chosen to check
eavesdropping. In this way, the protocol can be generalized to a
multiparty common key distribution protocol. Note that all of the
GCORE protocols have a final step, i.e. error correction and
privacy amplification [30], we shall not discuss these points,
which are the same as in all cryptographic protocols, except that
we have to use qutrits (qudits) instead of bits, and therefore
parity checks becomes triality checks, that is sums of modulo3
($d$).

In summary, we extend the idea of CORE to $N$-qubit,
$N$-qutrit quantum systems, propose the detailed protocols and
give the corresponding security analysis of 3-qubit, 2-qutrit
maximally entangled states, finally, we obtain the GCORE using the
general expression of multi-particle and high dimension maximally
entangled basis state by using repeatedly a {\it prior} shared
control key in this paper. The generalized version has great
capacity and high efficiency. In addition, the control key can be
used to control the GCORE operation of a group of units, so it
greatly simplifies the experimental realization and enables
quantum key distribution in a more efficient way.
\\
\\
\\
\\
\textbf{Acknowledgments} We thank Feng Xu, Ren-Gui Zhu, Xiao-Qiang
Su, Liang Qiu, De-Hui Zhan, Xue-Chao Li and Zhu-Qiang Zhang for
useful discussions. This project was supported by the National Basic
Research Programme of China under Grant No 2001CB309310, the
National Natural Science Foundation of China under Grant No
60573008.
\\
\\
\textbf{\textit{References:}}

[1] Bennett and G. Brassard, in Proceedings of the IEEE
International Conference on Computers, Systems and Signal
Processing, Bangalore, India $\sim $IEEE, New York, 1984,
pp.175--179.

[2]A. K. Ekert   Phys. Rev. Lett. 67, 661 (1991) .

[3]C.H. Bennett, G. Brassard, and N.D. Mermin   Phys. Rev. Lett.
68 557 (1992).

[4]C.H. Bennett, Phys. Rev. Lett. 68 3121 (1992).

[5]C.H. Bennett and S.J. Wiesner, Phys. Rev. Lett. 69 2881 (1992).

[6]W. Y. Hwang, I. G. Koh, and Y. D. Han, Phys. Lett. A 244 489
(1998).

[7]L.Goldenberg and L.Vaidman  Phys. Rev. Lett. 75 (1995) 1239;
Asher Peres  Phy. Rev. Lett. 77  3264 (1996); Goldenberg and
Vaidmann Phy. Rev. Lett. 77  3265 (1996).

[8]M.Koashi and N.Imoto Phys. Rev. Lett. 79  2383 (1999).

[9]Fu-Guo Deng and G. L. Long Phys. Rev. A 68  042315  (2003).

[10]H.Bechmann-Pasquinucci and A.Peres  Phys. Rev. Lett. 85 3313
(2000) .

[11]H.Bechmann-Pasquinucci and W.Tittel  Phys. Rev. A 61 062308
(2000).

[12]Mohamed Bourennane, Anders Karlsson and Gunnar Bjork Phys.
Rev. A 64 012306 (2001).

[13]Peng Xue, Chuan-Feng Li and Guang-Can Guo  Phy. Rev. A 64
032305 (2001).

[14]Peng Xue, Chuan-Feng Li and Guang-Can Guo  Phy. Rev. A 65
034302 (2002).

[15]K.Tamaki, M.Koashi and N.Imoto  Phys. Rev. Lett. 90 167904
(2003).

[16]A.Cabello Phys. Rev. Lett. 85 5635 (2000).

[17]G.L.Long and X.S.Liu  Phys. Rev. A 65  032302 (2002).

[18]N.J.Cerf, Mohamed Bourennane, Anders Karlsson and Nicolas
Gisin ~~~Phys. Rev. Lett. 88  127902  (2002).

[19]Nicolas J.Cerf, Thomas Durt and Nicolas Gisin   ~J.~Mod.~Opt.
49 1355 (2002), ~quant-ph/0110092.

[20]W. Dur and J. I. Cirac Phys. Rev. A 61 042314  (2000)
~~quant-ph/9911044.

[21] I.Csiszar and J.Korner, IEEE Trans.Inf. Theory 24, 339
(1978).

[22]M.Bourennane, A.Karlsson, G.Bjork, N.Gisin and N.J.Cerf ~~J.
Phys.A 35 10065 (2002) quant-ph/0106049.

[23]C.H.Bennett, G.Brassard, Claude Cr\'{e}peau, Richard Jozsa,
Asher Peres, and W. K. Wootters Phys. Rev. Lett. 70 1895  (1993).

[24]X.S.Liu, G.L.Long, D.M.Tong, and F. Li   Phys. Rev. A 65
022304 (2002).

[25]M. Zukowski, A. Zeilinger and M. A. Horne Phys. Rev. A 55 2564
(1997).

[26]Dagomir Kaszlikowski, L. C. Kwek, Jing-Ling Chen, Marek
Zukowski and C.H.Oh    Phys. Rev. A 65 032118  (2002).

[27]M.Zukowski, A.Zeilinger and M.A.Horne  Phys. Rev. A 55 2564
(1997).

[28]N.J.Cerf, Phys. Rev. Lett. 84  4497 (2000); J. Mod. Opt. 47
187 (2000); Acta Phas. Slovaca 48  115 (1998).

[29] C.H.Bennett, F.Bessette, G.Brassard, L.Salvail and J.Smolin
J. Cryptol. 5 3 (1992).

[30] P.D.Townsend, Nature (London) 385 47 (1997).

[31]S.J.D.Phoenix, S.M.Bnnett, P.D.Twensend and K.J.Blow J. Mod.
Opt. 42  1155  (1995).
\end{document}